UNIVERSITY
*of*
GLASGOW

Department of Physics & Astronomy
Experimental Particle Physics Group
Kelvin Building, University of Glasgow,
Glasgow, G12 8QQ, Scotland
Telephone: +44 (0)141 339 8855 Fax: +44 (0)141 334 9029

# MEASUREMENTS OF $\alpha_s$ FROM SCALING VIOLATIONS AND FROM R$_\tau$ AT LEP

Philip Reeves

**Abstract**

The results of two methods to extract the strong coupling constant, $\alpha_s$, are reviewed for the LEP experiments. In the first, scaling violations in the scaled momentum distributions of charged particles at LEP and at lower energy experiments are employed. In the second, QCD corrections to $R_\tau$, the ratio of hadronic to leptonic decay rates of the tau lepton, determine $\alpha_s$ and, using moments of the invariant mass distribution in hadronic tau decays, the non-perturbative corrections to $R_\tau$.

# 1 $\alpha_s$ from scaling violations

Study of the $Q^2$ dependence of structure functions in deep-inelastic scattering experiments provides a well-established method of measuring the strong coupling constant, $\alpha_s$. Analogously, $\alpha_s$ may be extracted from scaling violations in inclusive momentum distributions in the hadronic final state of $e^+e^-$ annihilation.

The general form for the inclusive distribution of hadrons in $x$ and polar angle $\theta$ with respect to the beam axis is given by[1)]:

$$\frac{d^2\sigma(s)}{dx\, d\cos\theta} = \frac{3}{8}(1+\cos^2\theta)\,\frac{d\sigma^T(s)}{dx} + \frac{3}{4}\sin^2\theta\,\frac{d\sigma^L(s)}{dx} + \frac{3}{4}\cos\theta\,\frac{d\sigma^A(s)}{dx}, \tag{1}$$

where $T$, $L$ and $A$ refer to the transverse, longitudinal and asymmetric cross sections. Integrating over $\cos\theta$ one obtains the inclusive cross section

$$\frac{d\sigma(s)}{dx} = \frac{d\sigma^T(s)}{dx} + \frac{d\sigma^L(s)}{dx} \tag{2}$$

The total cross section is dominated by the transverse component. The longitudinal component arises from QCD corrections and is used only to constrain the gluon fragmentation function. The cross sections are related to fragmentation functions $D_i$ ($i$ = u, d, s, c and b) for quarks and $D_g$ for gluons, which describe the momentum spectrum of final state particles from a single parton, by a convolution with coefficient functions[2)] $C_q$, $C_g$:

$$\begin{aligned} \frac{d\sigma(s)}{dx} &= 2\sigma_0(s)\int_x^1 \frac{dz}{z} C_q(z,\alpha_s(\mu_F),\mu_F^2/s) \sum_{i=u,d,s,c,b} w_i(s)\, D_i(x/z,\mu_F^2) \\ &+ 2\sigma_0(s)\int_x^1 \frac{dz}{z} C_g(z,\alpha_s(\mu_F),\mu_F^2/s)\, D_g(x/z,\mu_F^2)\ . \end{aligned} \tag{3}$$

Here $\sigma_0(s)$ is the Born cross section at the centre-of-mass energy $\sqrt{s}$ and $w_i$ is the relative electroweak cross section for the production of primary quarks of type $i$. The scale $\mu_F$ is an arbitrary factorization scale where the fragmentation functions are evaluated. The fragmentation functions themselves cannot be calculated within perturbative QCD, but once they are fixed at some parametrization scale $\sqrt{s_0}$, their energy evolution is predicted. The coefficient functions are known to next-to-leading order, $\mathcal{O}(\alpha_s)$. At leading order, only $C_q$ is different from zero, i.e. the cross section is proportional to the weighted sum of the quark fragmentation functions.

The QCD scaling violations are described by the Dokshitzer-Gribov-Lipatov-Altarelli-Parisi evolution equations[3)]

$$\frac{dD_j(x,s)}{d\ln s} = \sum_{i=u,d,s,c,b,g} \int_x^1 \frac{dz}{z} P_{ij}(z,\alpha_s(\mu_R),\mu_R^2/s)\, D_i(x/z,s)\ , \tag{4}$$

where $\mu_R$ is the renormalization scale. The splitting kernels[4,5)] $P_{ij}$ are known to next-to-leading order $\mathcal{O}(\alpha_s^2)$. Both the coefficient functions and the splitting kernels in the $\overline{\mathrm{MS}}$ scheme can be found, for example, in Nason and Webber[1)].

In principle, ratios of $x$ distributions measured at different energies differing from one would demonstrate the existence of scaling violations and allow a determination of $\alpha_s$. Differences, however, might be attributable to the different flavour composition from low energy PEP/PETRA data (mainly u-type quarks) to LEP data (mainly d-type quarks). In a recent analysis by the ALEPH collaboration, flavour-enriched samples have been used to fix the fragmentation functions of individual flavours in the data at a fixed $\sqrt{s}$, thus reducing the dependence on Monte Carlo models.

The fragmentation functions for each flavour were parametrized at a certain energy $\sqrt{s_0}$ as

$$D_i(x, s_0) = N_i x^{a_i} (1-x)^{b_i} e^{-c \ln^2(x)} \tag{5}$$

for $i =$ uds, c, b and g, motivated by the MLLA prediction of a Gaussian peak in $\ln(1/x)$ with a power law fall off as $x \to 1$. Non-perturbative effects in the evolution were parametrized as a shift in the effective value of $x$.

Measurements of the inclusive $x$ distributions of all flavours from PEP, PETRA, TRISTAN and DELPHI, spanning $\sqrt{s} = 22\,\mathrm{GeV}$ to $91.2\,\mathrm{GeV}$, were used. Along with the inclusive $x$ distribution, flavour-enriched samples of uds and b events were obtained by ALEPH with a lifetime (anti)tag, while gluon distributions were obtained from symmetric three-jet events and from the longitudinal cross-section (by weighting the polar angle distribution by[5] $(2 - 5\cos^2\theta)$).

With the fragmentation functions parametrized at $\sqrt{s_0} = 22\,\mathrm{GeV}$, $\alpha_s$ and the fragmentation parameters $N_i$, $a_i$, $b_i$ and $c$ were obtained in a simultaneous fit to all the data, evolving the fragmentation functions according to equation (4). Figure 1 shows the experimental distributions together with the QCD-evolved fit result. Only data in the range $0.1 < x < 0.8$ were used; outside this range the data are limited by systematic errors.

Normalization errors pose a problem for those experiments where only the combined statistical and systematic errors have been published. In those cases the purely statistical error was estimated from the amount of data that was available and subtracted from the total errors. Of the remaining relative errors in the respective $x$ bins, the minimum one was taken to be the normalization uncertainty common to all bins. This procedure was adopted for the nominal analysis. Alternatively all unspecified errors were taken as bin-to-bin errors and the resulting shift was taken as an additional systematic error.

The resulting value is $\alpha_s = 0.1265 \pm 0.0080\ (stat) \pm 0.0082\ (sys)$ where the systematic error arises from uncertainty in the factorization scale $\mu_F$ (0.0063), the renormalization scale $\mu_R$ (0.0020) and treatment of experimental errors quoted as a combined statistical and systematic error (0.0048). The $\chi^2/dof = 198/190$ shows that QCD evolution reproduces the observed scaling violations.

This analysis can be compared to a measurement by the DELPHI collaboration[6], in which inclusive $x$ distributions at PEP, PETRA and LEP were compared with the commplete $\mathcal{O}(\alpha_s^2)$ matrix element calculation, as implemented in JETSET. The fragmentation of quarks and gluons is modelled with the string mechanism. There are two fragmentation functions: the symmetric Lund function

$$f(z) \propto \frac{1}{z}(1-z)^a \tag{6}$$

for light quarks (u, d and s), and Peterson's function

$$f(z) \propto \frac{1}{z[1 - 1/z - \epsilon_b(1-z)]^2} \tag{7}$$

for heavy quarks (c and b).

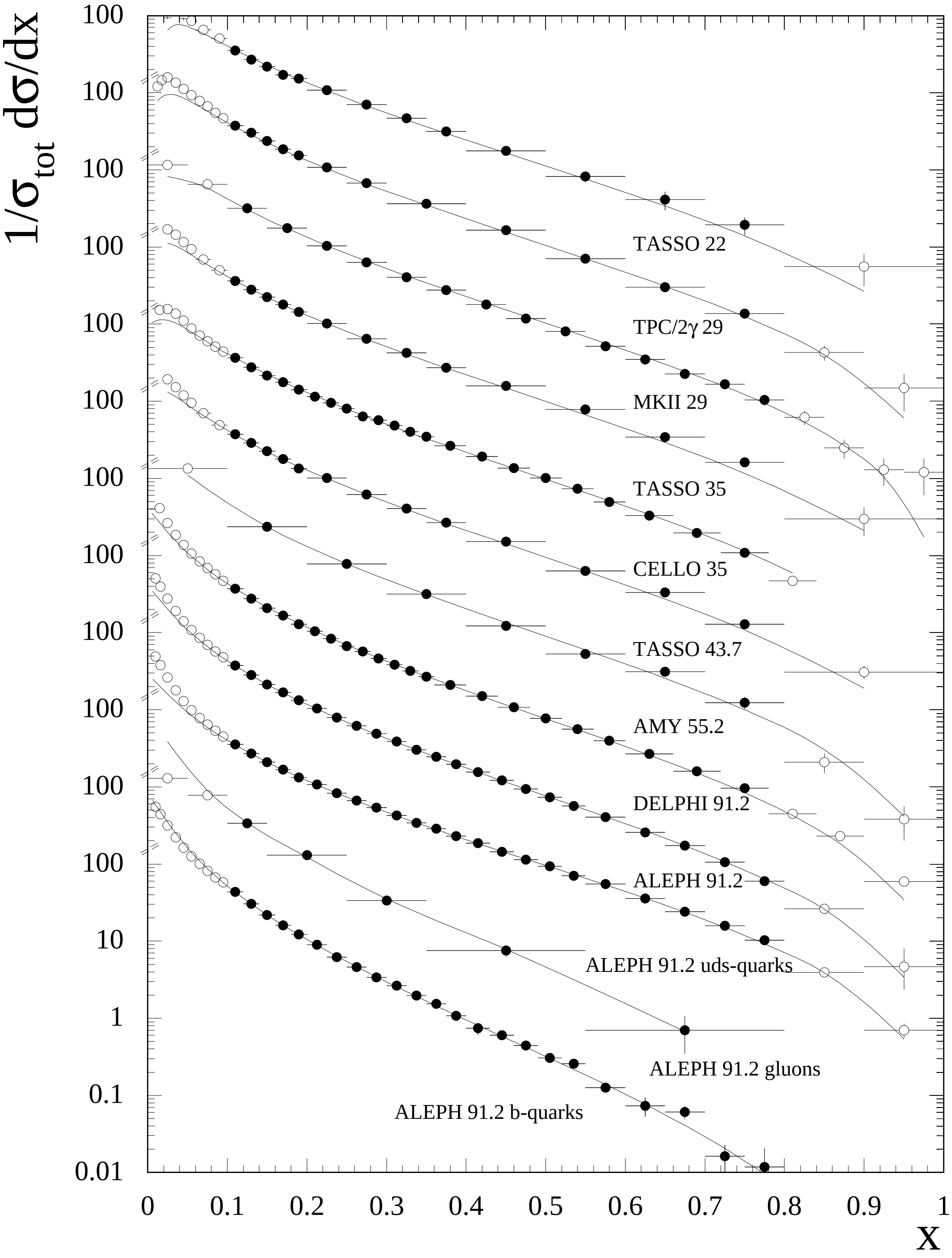

Figure 1: Inclusive $x$ distributions measured at various centre-of-mass energies. The line shows the fitted fragmentation functions evolved according to QCD to the energy of the experimental measurement. The open points were not used in the fit.

In a simultaneous fit for $\Lambda_{\overline{MS}}$, the light quark fragmentation parameter $a$ and the heavy quark fragmentation parameter $\epsilon_b$ the value $\alpha_s = 0.118 \pm 0.005$ was obtained, the error again dominated by scale uncertainties.

## 2 $\alpha_s$ from $R_\tau$

The quantity $R_\tau$

$$R_\tau = \frac{\Gamma(\tau^- \to \nu_\tau \text{ hadrons})}{\Gamma(\tau^- \to \nu_\tau \bar{\nu}_e e^-)} \approx N_c, \tag{8}$$

where $N_c$ is the number of colours, contains information on the strong coupling constant in the same way as does $R_Z$. The theoretical prediction[7,8,9)] for $R_\tau$ is obtained through an integral over the invariant mass-squared of the hadronic final state

$$R_\tau = \frac{1}{\Gamma_e} \int_0^{m_\tau^2} \frac{d\Gamma_h}{ds}\, ds. \tag{9}$$

Although $\Gamma_h$ is sensitive to non-perturbative effects and cannot be calculated, the behaviour of the integral is understood. In the Operator Product Expansion approach of Shifman, Vainshtein and Zakharov[10)] (SVZ), perturbative and non-perturbative contributions fall into a systematic expansion in $1/s$:

$$R_\tau = N_c(|V_{ud}|^2 + |V_{us}|^2) S_{EW}(1 + \delta_{EW} + \delta_{QCD} + \delta_m + \delta_{np}), \tag{10}$$

where $S_{EW}$ and $\delta_{EW}$ are purely electroweak corrections, $\delta_{QCD}$ is the perturbative QCD correction (known to $\mathcal{O}(\alpha_s^3)$ with a resummation to all orders[8)] of ill-behaved terms) and $\delta_m$ is a correction for the finite quark masses. The non-perturbative corrections are contained in $\delta_{np}$. In the SVZ approach, the $\mathcal{O}(2)$ and $\mathcal{O}(4)$ terms are zero. An $\mathcal{O}(4)$ contribution does arise, however, from the so-called gluon condensate, $\langle \frac{\alpha_s}{\pi} GG \rangle$. Numerical estimates from phenomenological fits give[7)] $\langle \frac{\alpha_s}{\pi} GG \rangle = 0.02 \pm 0.01\,\text{GeV}^4$ and $\mathcal{O}(6) = 0.002 \pm 0.001\,\text{GeV}^6$. The $\mathcal{O}(8)$ term was neglected. It must be emphasized that the reliability of the QCD prediction depends on the applicability of the SVZ approach, which has not yet been formally established.

Experimental measurements of $R_\tau$ are obtained from the leptonic branching ratios $R_\tau = (1 - B_e - B_\mu)/B_e$. An independent measurement is obtained from the tau lifetime using $B_e = (\tau_\tau/\tau_\mu)(m_\tau/m_\mu)^5$. In a recent analysis by the OPAL collaboration[11)] a value $R_\tau = 3.605 \pm 0.064$ was obtained from lepton branching ratios. Combining with $R_\tau = 3.682 \pm 0.048$ from the tau lifetime yields $R_\tau = 3.654 \pm 0.038$. Using the SVZ expansion with the non-perturbative coefficients given above, a value $\alpha_s(m_\tau) = 0.375^{+0.019+0.025+0.006}_{-0.018-0.017-0.006}$ was obtained. The first error is the combined experimental error, the second is due to unknown higher orders and the third is due to uncertainties in the non-perturbative contribution. Evolution to the $m_Z$ scale yields $\alpha_s(m_Z) = 0.1229^{+0.0016+0.0025}_{-0.0017-0.0021}$, where the first error is experimental and the second theoretical, including $\pm 0.0011$ from the evolution through c and b thresholds.

Given the integral (9), moments of the hadronic invariant mass distribution may be calculated:

$$R_\tau^{kl} = \frac{1}{\Gamma_e} \int_0^{m_\tau^2} ds \left(1 - \frac{s}{m_\tau^2}\right)^k \left(\frac{s}{m_\tau^2}\right)^l \frac{d\Gamma_h}{ds}, \tag{11}$$

$$D^{kl} = R^{kl}/R^{00}. \tag{12}$$

Similar predictions to (10) are known for the moments[9]. Using the moments $D^{10}$, $D^{11}$, $D^{12}$ and $D^{13}$, along with $R_\tau$ from lepton branching ratios and the tau lifetime, the ALEPH collaboration have performed a simultaneous fit for $\alpha_s$ and the non-perturbative coefficients $\langle \frac{\alpha_s}{\pi} GG \rangle$, $\mathcal{O}(6)$ and $\mathcal{O}(8)$.

The hadronic invariant mass distribution was obtained from the sum of those of several exclusively-reconstructed channels of up to five charged hadrons. Cabbibo-suppressed decays were subtracted according to the predictions of a Monte Carlo model and the distributions unfolded for detector efficiency and cross-talk between channels. The values of the moments are given in table 1.

Table 1: Moments of the hadronic invariant mass distribution in tau decays, measured by ALEPH.

| | moment | $\sigma_{stat}$ | $\sigma_{sys}$ |
|---|---|---|---|
| $D^{10}$ | 0.7217 | 0.0018 | 0.0060 |
| $D^{11}$ | 0.1556 | 0.0006 | 0.0018 |
| $D^{12}$ | 0.0570 | 0.0004 | 0.0012 |
| $D^{13}$ | 0.0259 | 0.0003 | 0.0007 |
| $R_\tau = 3.645 \pm 0.024$ | | | |

In a fit to the moments and $R_\tau$, a value $\alpha_s(m_\tau) = 0.355 \pm 0.021$ was obtained. The non-perturbative coefficients were $\langle \frac{\alpha_s}{\pi} GG \rangle = 0.0063 \pm 0.012\,\mathrm{GeV}^4$, $\mathcal{O}(6) = -0.0016 \pm 0.0015\,\mathrm{GeV}^6$ and $\mathcal{O}(8) = 0.0021 \pm 0.0018\,\mathrm{GeV}^8$, which may be compared with the values[7] given above. Evolution to the $m_Z$ scale yields $\alpha_s(m_Z) = 0.121 \pm 0.0022 \pm 0.0010$, where the second error is from the evolution through c and b thresholds.

# 3 Summary

Using flavour-enriched samples to fix the quark and gluon fragmentation functions, the strong coupling constant $\alpha_s$ has been determined by the ALEPH collaboration from scaling violations in the inclusive $x$ distributions of hadrons, yielding $\alpha_s(m_Z) = 0.127 \pm 0.011$. Two measurements of $\alpha_s$ from $R_\tau$ have been presented. Determining $R_\tau$ from lepton branching ratios and the tau lifetime, the OPAL collaboration have obtained $\alpha_s(m_\tau) = 0.375^{+0.032}_{-0.025}$, or $\alpha_s(m_Z) = 0.1229^{+0.0030}_{-0.0027}$ evolved to the $m_Z$ scale. Using moments of the invariant mass in hadronic tau decays, the ALEPH collaboration have simultaneously fitted $\alpha_s$ and the non-perturbative corrections to $R_\tau$: $\alpha_s(m_\tau) = 0.355 \pm 0.021$, $\langle \frac{\alpha_s}{\pi} GG \rangle = 0.0063 \pm 0.012\,\mathrm{GeV}^4$, $\mathcal{O}(6) = -0.0016 \pm 0.0015\,\mathrm{GeV}^6$ and $\mathcal{O}(8) = 0.0021 \pm 0.0018\,\mathrm{GeV}^8$. Evolving to the $m_Z$ scale, $\alpha_s(m_Z) = 0.121 \pm 0.0024$. The results are in good agreement with other determinations of $\alpha_s$ at LEP (from the Z lineshape[12], for example).